\documentclass[preprint,aps]{revtex4-1}
\usepackage{graphicx}
\usepackage{hyperref}
\usepackage{epstopdf}
\usepackage{slashed}
\usepackage{rotating}
\usepackage{float}
\newcommand{\be}{\begin{equation}}
\newcommand{\ee}{\end{equation}}
\newcommand{\bea}{\begin{eqnarray}}
\newcommand{\eea}{\end{eqnarray}}
\newcommand{\ba}{\begin{array}}
\newcommand{\ea}{\end{array}}

\newcommand{\TL}[6]{#1 & #2 & #3 & #4& #5& #6  \\}

\begin{document}
\title{Quark flavor distribution functions for the octet baryons in the chiral quark constituent model}
\author{Harleen Dahiya}
\affiliation{Department of Physics,\\ Dr. B.R. Ambedkar National
Institute of Technology,\\ Jalandhar, 144011, India}
\begin{abstract}
The quark flavor distribution functions of the octet baryons ($N$, $\Sigma$, $\Xi$ and $\Lambda$) have been calculated in
the chiral constituent quark model ($\chi$CQM).
In particular,  the valence and sea quark flavor distribution functions of the scalar density matrix elements of octet baryons have been computed explicitly. The implications of chiral symmetry breaking and SU(3) symmetry
breaking have been discussed in detail for the sea quark asymmetries, fraction of a particular quark (antiquark) present in a baryon, flavor structure functions and the Gottfried integral. The meson-baryon sigma terms $\sigma_{\pi B}$, $\sigma_{K B}$, and $\sigma_{\eta B}$ for the case of $N$, $\Sigma$ and $\Xi$ baryons have also been calculated. The results  have been compared with the recent available experimental observations for the case of $N$ and how the  future experiments for $\Sigma$, $\Xi$ and $\Lambda$ can provide important constraints to describe the role of non-valence (sea) degrees of freedom has been discussed.

\end{abstract}

\maketitle
\section{Introduction}
The fact that the quarks are point-like constituents was revealed in the deep
inelastic scattering (DIS) \cite{point-like}.  These point-like constituents  were identified as the valence or constituent quarks
with spin-$\frac{1}{2}$ in the naive quark model (NQM)
\cite{dgg,Isgur,yaouanc,mgupta}.  This was further confirmed by the measurements of polarized structure functions of proton, to have a deeper insight into the  internal structure of the baryons, in the DIS experiments \cite{emc,smc,adams,hermes_spin}. Surprisingly, these DIS results provided the first evidence that the valence quarks of proton carry only about 30\% of its spin clearly indicating that they should be surrounded by an indistinct sea of quark-antiquark pairs.
Recently, experiments measuring the weak and electromagnetic form factors from the elastic scattering of electrons have provided considerable insight on the role played by strange quarks in the charge, current and spin structure of the nucleon.  For
example, SAMPLE at MIT-Bates \cite{sample}, G0 at JLab \cite{g0},
PVA4 at MAMI \cite{a4} and HAPPEX  at JLab \cite{happex} have provided considerable insight on the role played
by strange quarks when the nucleon interacts at high energies and have clearly indicated explicitly the non-valence contribution in the nucleon which is otherwise absent in the NQM picture.
Even though the internal structure of the nucleon has been extensively studied over the past 40 or 50 years but because of confinement, the knowledge has been rather limited and it is still a big challenge to perform the calculations from the first principles of Quantum Chromodynamics (QCD) and as a result the study of the composition of hadrons still remains to be a major unresolved issue in high energy spin physics.

Apart from the spin structure, several interesting facts have also been revealed regarding the
flavor structure of the sea quark content in the nucleon. Major surprise was found when the famous DIS experiments by the New Muon
Collaboration (NMC) in 1991 \cite{nmc} established the sea quark
asymmetry of the unpolarized quarks in the case of nucleon by
measuring the violation of the Gottfried sum rule (GSR) ($\int_0^1[\bar d (x)-\bar u (x)]dx$) \cite{gsr}. This was subsequently confirmed by two independent experiments in various $0 \leq x \leq 1$ ranges. First from the Fermilab E866 experiments \cite{e866}, measuring a large sea quark  asymmetry ratio $\bar d/\bar u$ as well as $\bar d-\bar u \neq 0$, and the other from the Drell-Yan cross section ratios of the NA51 experiments \cite{baldit}.  More recently, HERMES has presented
another $u-d$ sea quark asymmetry $\frac{\bar d-\bar u}{u-d}$ \cite{hermes_flavor} confirming the violation of GSR. There was a clear indication from these results that the
structure of the nucleon is not limited to $u$ and $d$
quarks and the origin of the sea quarks should be nonperturbative in nature because the conventional expectation that the sea quarks  perhaps can be
obtained through the perturbative production of the
quark-antiquark pairs by gluons producing nearly equal numbers of
$\bar u$ and $\bar d$.

In addition, the information on the strange sea is obtained from the neutrino-induced DIS experiments \cite{neudis} as well as through the charm production with dimuon events in the final states of the experiments CDHS \cite{cdhs}, CCFR \cite{ccfr1,ccfr2}, CHARMII \cite{charmii}, NOMAD \cite{nomad1,nomad2}, NuTeV \cite{nutev} and CHORUS \cite{chorus}. It has been emphasized in the neutrino-induced DIS experiments that the valence quark
distributions dominate for $x > 0.3$ and it is a relatively clean
region to test the valence structure of the nucleon as well as to
estimate the structure functions and related quantities, whereas the sea quarks dominate for the $x<0.3$. These experiments  have renewed considerable interest in the sea quark flavor structure as well as asymmetries and they point out the need for additional refined data. In this regard, the ongoing Drell-Yan experiment at Fermilab \cite{fermilab} and a proposed experiment at J-PARC facility \cite{jparc} are working towards extending the kinematic coverage and improving the accuracy of the sea quark  asymmetry.

In the context of low-energy experiments \cite{nutev,pion nucleon}, the pion-nucleon sigma term ($\sigma_{\pi N}$) has received much attention in the past. It has been determined precisely from the pion-nucleon scattering experiments \cite{koch,gasser,Pavan:2001wz,hite} as well as hadron spectroscopy \cite{riaz}. The results from both the methods however differ substantially.
The $\sigma_{\pi N}$ term is known to have intimate connection with the dynamics of the non-valence quarks and is an important fundamental parameter to test the chiral symmetry breaking effects and thereby determine the scalar quark content of the
nucleon. In addition it also provides restriction on the contribution of strangeness to the parameters measured in low-energy \cite{bass} since the strange quarks constitute purely sea degrees of freedom. Our experimental information about the other meson-baryon sigma terms $\sigma_{\pi B}$, $\sigma_{K B}$, and $\sigma_{\eta B}$, for the case of $N$, $\Sigma$, $\Xi$ and $\Lambda$ baryons, is also rather limited because of the difficulty in the measurements due to their short lifetimes. The low-energy determination of $\sigma_{M B}$ would undoubtedly provide
vital clues to the nonperturbative aspects of QCD.

Even though there has been considerable progress in the past few years to understand the origin of the sea quark flavor structure, there is no consensus regarding the various mechanisms which can contribute to it. This has motivated a large amount of effort to understand the origins of the nucleon sea. The broader question of non-valence quark contribution to the unpolarized distributions of sea quarks, sea quark asymmetry, structure function has been discussed \cite{ellis-brodsky,alkofer,christov,diakonov,mesoncloud,wakamatsu,eccm,stat,bag,alwall,reya,chang-14}. One of the most successful nonperturbative approach is the chiral constituent
quark model ($\chi$CQM) \cite{manohar}.  The basic idea is based on
the possibility that chiral symmetry breaking takes place at
a distance scale much smaller than the confinement scale.
The $\chi$CQM uses the effective interaction Lagrangian
approach of the strong interactions  where the effective degrees of freedom are the
valence quarks and the internal Goldstone bosons (GBs)
which are coupled to the valence quarks \cite{cheng,johan,song,hd}. The $\chi$CQM successfully explains  the ``proton
spin problem'' \cite{hd}, magnetic moments of
octet and decuplet baryons including their
transitions and the Coleman-Glashow sum rule \cite{hdmagnetic}, hyperon $\beta$ decay parameters
\cite{nsweak}, magnetic moments of octet
baryon resonances \cite{nres}, magnetic moments of $\Lambda$ resonances \cite{torres}, charge radii and quadrupole moment \cite{charge-radii}, etc..
The model is successfully extended to
predict the important role played by the small intrinsic charm content in the nucleon spin in the SU(4) $\chi$CQM  and to
calculate the magnetic moment and charge radii of charm baryons including
their radiative decays \cite{hdcharm}. In view of the above developments in the $\chi$CQM, it become desirable to extend the model to calculate the quark flavor distribution functions and related quantities of the octet baryons whose knowledge would undoubtedly provide vital clues to the nonperturbative aspects of QCD.

The purpose of the present communication is to determine the quark flavor distribution functions of the octet baryons in
the chiral constituent quark model ($\chi$CQM) which is one of the most successful models to phenomenologically estimate the quantities affected
by chiral symmetry breaking and SU(3) symmetry breaking. In particular, we would like to understand in detail the implications of the scalar density matrix elements of octet baryons in terms of the valence and sea quark flavor distribution functions, related sea quark  asymmetries, fractions of quarks and antiquarks present in a baryon, flavor structure functions and the Gottfried integral.
Further, it would be interesting to extend the calculations to predict the meson-baryon sigma terms $\sigma_{\pi B}$, $\sigma_{K B}$, and $\sigma_{\eta B}$ for the case of $N$, $\Sigma$, $\Xi$ and $\Lambda$ baryons. The results can be compared with the recent available experimental observations and can also provide important constraints on the future experiments to describe the role of non-valence degrees of freedom.

\section{Chiral Constituent Quark Model}
The $\chi$CQM was
introduced by Weinberg and further developed by
Manohar and Georgi \cite{manohar}.
The underlying idea is that the set of internal
Goldstone bosons (GBs) couple directly to the valence quarks in the
interior of hadron, but not at so small distances that perturbative
QCD is applicable.

The dynamics of light quarks ($u$, $d$, and $s$) and
gluons can be described by the QCD Lagrangian
\be
{\cal{L}} = - \frac{1}{4}G_{\mu \nu}^{a} G^{\mu \nu}_{a} + i
\bar{\psi}_R \slashed{D} {\psi}_R + i \bar{\psi}_L \slashed{D}
{\psi}_L - \bar{\psi}_R M {\psi}_L - \bar{\psi}_L M {\psi}_R \,,
\label{lagrang1} \ee where $ G_{\mu \nu}^{a}$ is the gluonic gauge field strength tensor,
$D^{\mu}$ is the gauge-covariant derivative,  $M$ is the quark mass matrix and $\psi_L$  and $\psi_R$ are the left and right
handed quark fields respectively
\be \Psi_L \equiv \left(\ba{c}
u_L\\d_L\\s_L \ea \right) ~~~ {\rm and } ~~~ \Psi_R \equiv \left(\ba{c}
u_R\\d_R\\s_R \ea \right)\,. \ee

Since the mass
terms change sign  as $\psi_{R}
\to \psi_{R}$ and $\psi_{L} \to -\psi_{L}$ under the chiral transformation $(\psi \to \gamma^5 \psi)$, the Lagrangian in Eq. (\ref{lagrang1}) no
longer remains invariant. In case the mass terms in the QCD Lagrangian
are neglected, the Lagrangian will have global chiral symmetry of
the {\it SU}(3)$_L$$\times${\it SU}(3)$_R$ group. Since the
spectrum of hadrons in the known sector does not display parity
doublets, the chiral symmetry is believed to be spontaneously broken
around a scale of 1 GeV as \be {\it SU}(3)_L \times
{\it SU}(3)_R \to {\it SU}(3)_{L+R} \,. \ee As a consequence, there
exists a set of massless particles, referred to as the Goldstone
bosons (GBs), which are identified with the observed ($\pi$, $K$,
$\eta$ mesons). Within the region of QCD confinement scale
($\Lambda_{QCD} \simeq 0.1-0.3$ GeV) and the chiral symmetry breaking scale
$\Lambda_{\chi SB}$, the constituent quarks, the octet of GBs
($\pi$, K, $\eta$ mesons), and the {\it weakly} interacting gluons
are the appropriate degrees of freedom.

The effective interaction Lagrangian in this region can be expressed as \be {\cal
L}_{{\rm int}} = \bar{\psi}(i{\slashed D} + {\slashed V})\psi +
ig_{A} \bar{\psi} {\slashed A}\gamma^{5}\psi + \cdots \,,
\label{lagrang2} \ee where $g_{A}$ is the axial-vector coupling
constant. The gluonic degrees of
freedom can be neglected owing to small effect in the
effective quark model at low energy scale. The vector and axial-vector currents
$V_{\mu}$ and $A_{\mu}$ are defined as \be \left( \ba{c}
  V_{\mu} \\
  A_{\mu} \\ \ea
\right)=\frac{1}{2}(\xi^{\dagger}\partial_{\mu}
\xi\pm\xi\partial_{\mu}\xi^{\dagger}),\ee where $\xi=\mathrm{exp}( 2
i \Phi/f_{\pi})$, $f_{\pi}$ is the pseudoscalar pion decay constant
($\simeq 93$~MeV), and $\Phi$ is the field describing the dynamics
of GBs as \bea \Phi = \left( \ba{ccc} \frac{\pi^0}{\sqrt 2} +
\beta\frac{\eta}{\sqrt 6} & \pi^+ & \alpha K^+ \\ \pi^- &
-\frac{\pi^0}{\sqrt 2} + \beta \frac{\eta}{\sqrt 6} & \alpha K^0 \\
\alpha K^- & \alpha \bar{K}^0 & -\beta \frac{2\eta}{\sqrt 6} \ea
\right)\,. \eea
Expanding $V_{\mu}$ and $A_{\mu}$ in the powers of
$\Phi/f_{\pi}$, we get \bea V_{\mu} &=& 0 +
O \left( (\Phi/f_{\pi})^{2} \right) \,, \\
A_{\mu} &=& \frac{i}{f_{\pi}} \partial_{\mu} \Phi + O \left(
(\Phi/f_{\pi})^{2} \right)\,.\eea

The effective interaction Lagrangian  between
GBs and quarks from Eq. (\ref{lagrang2}) in the leading order can now be expressed as \be {\cal
L}_{{\rm int}} = -\frac{g_{A}}{f_{\pi}} \bar{\psi} \partial_{\mu}
\Phi \gamma^{\mu} \gamma^{5} \psi \,, \label{lagrang3} \ee which can
be reduced to \be {\cal L}_{{\rm int}} \approx  i \sum_{q = u,d,s}
\frac{m_q + m_{q'}}{f_{\pi}} {\bar q'} \Phi \gamma^5 q = i
\sum_{q=u,d,s} c_8 {\bar q'} \Phi \gamma^5 q \,, \ee using the Dirac
equation $(i \gamma^{\mu} \partial_{\mu} - m_q)q =0$. Here,
$c_8\left( = \frac{m_q + m_{q'}}{f_{\pi}} \right)$ is the coupling
constant for octet of GBs and $m_q$ ($m_{q'}$) is the quark mass
parameter. The Lagrangian of the quark-GB interaction, suppressing
all the space-time structure to the lowest order, can now be
expressed as \be {\cal L}_{{\rm int}} =  c_8 {\bar \psi} \Phi \psi
\,.\ee

The QCD Lagrangian is also invariant under the axial
$U(1)$ symmetry, which would imply the existence of ninth GB. This
breaking symmetry picks the $\eta'$ as the ninth GB. The effective
Lagrangian describing interaction between quarks and a nonet of GBs,
consisting of octet and a singlet, can now be expressed as \be {\cal
L}_{{\rm int}} = c_8 { \bar \psi} \Phi {\psi} + c_1{ \bar \psi}
\frac{\eta'}{\sqrt 3}{\psi}= c_8 {\bar \psi}\left( \Phi + \zeta
\frac{\eta'}{\sqrt 3}I \right) {\psi }=c_8 {\bar \psi} \left(\Phi'
\right) {\psi} \,, \label{lagrang4} \ee where $\zeta=c_1/c_8$, $c_1$
is the coupling constant for the singlet GB and $I$ is the $3\times 3$
identity matrix.

The fluctuation process describing the effective Lagrangian is
\be q^{\pm} \rightarrow {\rm GB} + q^{'
\mp} \rightarrow (q \bar q^{'}) +q^{'\mp}\,, \label{basic}
\ee
where $q \bar q^{'} +q^{'}$
constitute the sea quarks \cite{cheng,johan,hd}. The GB field can be expressed in terms of the GBs and their transition probabilities as \bea
\Phi' = \left( \ba{ccc} \frac{\pi^0}{\sqrt 2}
+\beta\frac{\eta}{\sqrt 6}+\zeta\frac{\eta^{'}}{\sqrt 3} & \pi^+
  & \alpha K^+   \\
\pi^- & -\frac{\pi^0}{\sqrt 2} +\beta \frac{\eta}{\sqrt 6}
+\zeta\frac{\eta^{'}}{\sqrt 3}  &  \alpha K^o  \\
 \alpha K^-  &  \alpha \bar{K}^0  &  -\beta \frac{2\eta}{\sqrt 6}
 +\zeta\frac{\eta^{'}}{\sqrt 3} \ea \right). \eea
The transition probability of chiral
fluctuation  $u(d) \rightarrow d(u) + \pi^{+(-)}$, given in terms of the coupling constant for the octet GBs $|c_8|^2$, is defined as $a$ and is introduced by considering nondegenerate quark masses $M_s > M_{u,d}$. In terms of $a$, the probabilities of transitions of $u(d) \rightarrow s + K^{+(0)}$, $u(d,s)\rightarrow u(d,s) + \eta$, and $u(d,s) \rightarrow u(d,s) + \eta^{'}$ are given as $\alpha^2 a$, $\beta^2 a$ and $\zeta^2 a$ respectively \cite{cheng,johan}. The parameters $\alpha$ and $\beta$ are introduced by considering nondegenerate GB masses $M_{K},M_{\eta}> M_{\pi}$ and the parameter $\zeta$ is introduced by considering  $M_{\eta^{'}} > M_{K},M_{\eta}$. These parameters provide the basis to
understand the extent to which the sea quarks contribute to the
structure of the baryon. The hierarchy for the probabilities, which scale as $\frac{1}{M_q^2}$, can be obtained as
\be a> a \alpha^2 \geq a \beta^2> a \zeta^2.
\ee

The sea quark flavor distribution functions can be calculated in $\chi$CQM by substituting for every
valence (constituent) quark \be q \to  P_q q + |\psi(q)|^2, \label{substitute}\ee
where the transition probability of no emission of
GB $P_q$ can be expressed in terms of the transition probability of the emission of a GB from
any of the $u$, $d$ $s$ quark as follows
\be
P_q=1-P_{[q, ~GB]}, \label{probability} \ee with \be P_{[u, ~GB]}=P_{[d, ~GB]}
=a\left(\frac{3}{2}+\alpha^2+\frac{\beta^2}{6}+\frac{\zeta^2}{3}\right)\,,~~~~{\rm and}~~~~
P_{[s, ~GB]} = a\left(2 \alpha^2+\frac{2\beta^2}{3}+\frac{\zeta^2}{3} \right)\,, \label{probuds} \ee
whereas $|\psi(q)|^2$  is the transition
probability of the $q$  quark calculated from the Lagrangian expressed as

\bea |\psi(u)|^2 &=& a\left[\frac{7}{4} +\frac{\beta}{6} +\frac{\zeta}{3}+\frac{\beta \zeta}{9} +\alpha^2+\frac{7\beta^2}{36}+\frac{4\zeta^2}{9}\right]{ u} +\left[\frac{1}{4} +\frac{\beta}{6} +\frac{\zeta}{3}+\frac{\beta \zeta}{9} +\frac{\beta^2}{36}+\frac{\zeta^2}{9}\right]{ {\bar u}}\nonumber\\ &+& \left[\frac{5}{4} -\frac{\beta}{6} -\frac{\zeta}{3}+\frac{\beta \zeta}{9} +\frac{\beta^2}{36}+\frac{\zeta^2}{9}\right]({d}+{ {\bar d}})+ \left[-\frac{2\beta \zeta}{9} +\alpha^2+\frac{\beta^2}{9}+\frac{\zeta^2}{9}\right]({ s}+{ {\bar s}})\,,\label{eqpsiu}\eea
\bea |\psi(d)|^2 &=& a\left[\frac{7}{4} +\frac{\beta}{6} +\frac{\zeta}{3}+\frac{\beta \zeta}{9} +\alpha^2+\frac{7\beta^2}{36}+\frac{4\zeta^2}{9}\right]{ d} +\left[\frac{1}{4} +\frac{\beta}{6} +\frac{\zeta}{3}+\frac{\beta \zeta}{9} +\frac{\beta^2}{36}+\frac{\zeta^2}{9}\right]{ {\bar d}}\nonumber\\ &+& \left[\frac{5}{4} -\frac{\beta}{6} -\frac{\zeta}{3}+\frac{\beta \zeta}{9} +\frac{\beta^2}{36}+\frac{\zeta^2}{9}\right]({u}+{ {\bar u}})+ \left[-\frac{2\beta \zeta}{9} +\alpha^2+\frac{\beta^2}{9}+\frac{\zeta^2}{9}\right]({ s}+{ {\bar s}})\,,\label{eqpsid}\eea
\bea |\psi(s)|^2 &=& a\left[\frac{4\beta \zeta}{9} +2\alpha^2 +\frac{10\beta^2}{9}+\frac{4\zeta^2}{9}\right]{ s} +\left[\frac{4\beta \zeta}{9} +\frac{4\beta^2}{9}+\frac{\zeta^2}{9}\right]{ {\bar s}}\nonumber\\ &+& \left[-\frac{2\beta \zeta}{9} +\alpha^2 +\frac{\beta^2}{9}+\frac{\zeta^2}{9}\right]({u}+{ {\bar u}}+{ d}+{ {\bar d}})\,.\label{eqpsis}\eea

The flavor structure for the baryon
of the type $B(q_1q_2q_3)$ for the case of octet baryons having $q_1,q_2,q_3=u,d,s$ is expressed as \be P_{q_1} q_1 + P_{q_2} q_2 + P_{q_3} q_1+ |\psi(q_1)|^2 + |\psi(q_2)|^2+
|\psi(q_3)|^2.\ee

\section{quark flavor distribution functions}

The quark flavor distribution functions can be calculated from the scalar matrix elements of the octet baryons and can be defined as follows \cite{cheng}
\be \hat B \equiv \langle B|{\cal N}_{q \bar q}|B \rangle, \label{bnb}
\ee
where $|B\rangle$ is the
SU(6) baryon wavefunction (detailed in Ref. \cite{yaoubook}) and ${\cal N}_{q \bar q}$ is the number operator measuring the sum of the quark and antiquark numbers
\bea {\cal N}_{q \bar q}=
\sum_{q=u,d,s} (n_q q + n_{\bar q }{\bar q})&=& n_{u}u + n_{{\bar
u}}{\bar u} + n_{d}d + n_{{\bar d}}{\bar d} + n_{s}s + n_{{\bar
s}}{\bar s} \nonumber \\ &=& (n_{u}- n_{{\bar
u}})u + (n_{d}- n_{{\bar d}})d + (n_{s}- n_{{\bar
s}})s\,, \label{number} \eea
with the coefficients $n_{q({\bar q})}$ being the
number of $q({\bar q})$ quarks with electric charge $e_q(e_{\bar q}$). We have also used $q = -{\bar q}$ for a given baryon in the above equation.

The quark flavor distribution functions of the baryon receive contribution from the valence as well as the  sea quark distribution functions as follows
\be q^B=q^B_{{\rm V}}+q^B_{{\rm S}}.\ee
Since the antiquark distribution functions come purely from the sea quarks  therefore we can replace the sea quark distribution functions with the antiquark distribution functions as
\be q^B=q^B_{{\rm V}}+\bar q^B. \label{totalquark}\ee

The normalization conditions for the valence quark distribution functions of the octet baryons can be summarized in Table \ref{valencequark}. The antiquark densities of the octet baryons $p$, $n$, $\Sigma^+$, $\Sigma^-$,
$\Xi^0$, $\Xi^-$ and $\Lambda^0$ can easily be calculated using Eqs. (\ref{substitute}), (\ref{probability}), (\ref{eqpsiu}),
(\ref{eqpsid}) and (\ref{eqpsis}). The results have been presented in Table \ref{antiquark}.

\begin{table}[h]
\begin{tabular}{|c|c|c|c|}
\hline Baryon & $\int_0^1 u_{\rm V}^B(x)dx$ & $\int_0^1 d_{\rm V}^B(x)dx$ & $\int_0^1 s_{\rm V}^B(x)dx$ \\ \hline

$p(uud)$ & $2$ & $1$ & $0$ \\

$n(udd)$ & $1$ & $2$ & $0$ \\

$\Sigma^{+}(uus)$ & $2$ & $0$ & $1$ \\

$\Sigma^{-}(dds)$ & $0$ & $2$ & $1$ \\

$\Xi^{0}(uss)$ & $1$ & $0$ & $2$ \\

$\Xi^{-}(dss)$ & $2$ & $1$ & $0$ \\

$\Lambda^{0}(uds)$ & $1$ & $1$ & $1$ \\

\hline
\end{tabular}
\caption{The normalization conditions for the valence quark distribution functions of the octet
baryons integrated over the Bjorken variable $x$.}\label{valencequark}
\end{table}

\begin{table}
\begin{sideways}
\begin{tabular}{|c|c|c|c|}
\hline Baryon & $\bar u^B$ & $\bar d^B$ & $\bar s^B$ \\ \hline

$p(uud)$ & $\frac{a}{12}\left(21+\beta^2+4 \zeta+4\zeta^2+\beta(2+4 \zeta)\right)$ &
$\frac{a}{12}\left(33+\beta^2-4 \zeta+4\zeta^2+\beta(-2+4 \zeta)\right)$  & $3a\left(\alpha^2+\frac{1}{9}(\beta- \zeta)^2\right)$ \\

$n(udd)$ & $\frac{a}{12}\left(33+\beta^2-4 \zeta+4\zeta^2+\beta(-2+4 \zeta)\right)$  & $\frac{a}{12}\left(21+\beta^2+4 \zeta+4\zeta^2+\beta(2+4 \zeta)\right)$ & $3a\left(\alpha^2+\frac{1}{9}(\beta- \zeta)^2\right)$\\

$\Sigma^{+}(uus)$ & $\frac{a}{6}\left(3+6\alpha^2+ 2\beta+\beta^2+4 \zeta+2 \zeta^2 \right)$ & $\frac{a}{6}\left(15+6\alpha^2- 2\beta+\beta^2-4 \zeta+2 \zeta^2 \right)$ &$\frac{a}{3}\left(6\alpha^2+ 2\beta^2+ \zeta^2 \right)$ \\

$\Sigma^{-}(dds)$ & $\frac{a}{6}\left(15+6\alpha^2- 2\beta+\beta^2-4 \zeta+2 \zeta^2 \right)$ & $\frac{a}{6}\left(3+6\alpha^2+ 2\beta+\beta^2+4 \zeta+2 \zeta^2 \right)$
& $\frac{a}{3}\left(6\alpha^2+ 2\beta^2+ \zeta^2 \right)$ \\


$\Xi^{0}(uss)$ & $a \left( 2 \left(\alpha^2+\frac{1}{9} (\beta-\zeta)^2\right)+ \frac{1}{36} (3+\beta+2 \zeta)^2\right)$ & $a \left( 1+2 \left(\alpha^2+\frac{1}{9} (\beta-\zeta)^2\right)+ \frac{1}{36} (-3+\beta+2 \zeta)^2\right)$ & $\frac{a}{3}\left( 3 \alpha^2+3 \beta^2+2 \beta \zeta +\zeta^2\right)$ \\

$\Xi^{-}(dss)$ & $a \left( 1+2 \left(\alpha^2+\frac{1}{9} (\beta-\zeta)^2\right)+ \frac{1}{36} (-3+\beta+2 \zeta)^2\right)$& $a \left( 2 \left(\alpha^2+\frac{1}{9} (\beta-\zeta)^2\right)+ \frac{1}{36} (3+\beta+2 \zeta)^2\right)$ & $\frac{a}{3}\left( 3 \alpha^2+3 \beta^2+2 \beta \zeta +\zeta^2\right)$  \\

$\Lambda^{0}(uds)$ & $\frac{a}{6}\left( 9+6 \alpha^2+\beta^2+2\zeta^2\right)$ & $\frac{a}{6}\left( 9+6 \alpha^2+\beta^2+2\zeta^2\right)$ & $\frac{a}{3}\left( 6 \alpha^2+2 \beta^2+\zeta^2\right)$ \\

\hline
\end{tabular}
\end{sideways}
\caption{The sea quark (antiquark) distribution functions for the octet
baryons.}\label{antiquark}
\end{table}

In order to study the flavor structure of the baryons, we can
define the fraction of
particular quark and antiquark present in a baryon relative to the total number
of the quarks and antiquarks as
\be
f_q^B=\frac{q^B+\bar q^B}{\sum (q^B+\bar q^B)}\,, \label{fractionq}
\ee
where $q^B$ and $\bar q^B$ are the number of quarks and antiquarks for the octet baryon $B$ and $\sum (q^B+\bar q^B)$ is the sum of all the
quarks and antiquarks present.

Further, we can define
\bea
f_0^B&=&f_u^B+f_d^B+f_s^B,\nonumber \\
f_3^B&=&f_u^B-f_d^B,\nonumber \\
f_8^B&=&f_u^B+f_d^B-2 f_s^B.
\eea

Another relevant quantities are the suppression factors ($\rho^B$ and $\kappa^B$) of the strange quark
content with respect to the non-strange quarks and sea quarks
\bea
\rho_s^B&=&\frac{ s^B+\bar{s}^B}{ u^B+ d^B},\nonumber\\
\kappa_s^B&=&\frac{  s^B+ \bar{s}^B}{\bar{u}^B+ \bar{d}^B},
\eea
and the ratio of total number of the antiquarks and quarks
\be\frac{\sum \bar q^B}{\sum q^B}.
\ee

In order to calculate the phenomenological quantities pertaining to the valence and sea quark flavor distribution
functions, we first fix $\chi$CQM parameters pertaining to the probabilities of fluctuations to
pions, $K$, $\eta$, $\eta^{'}$) coming in the sea quark
distribution functions by taking into account strong physical considerations and carrying out a fine grained analysis using the well known experimentally measurable quantities pertaining to the spin and flavor distribution functions. The input
parameters and their values have been summarized in Table \ref{input}.

\begin{table}
\begin{center}
\begin{tabular}{|c|c|c|c|c|c|}      \hline
Parameter$\rightarrow$ &$a$ &$\alpha$ &$\beta$ &$\zeta$ & $\frac{m_s}{\hat m}$ \\
\hline
Value &  0.114 & 0.45 & 0.45 & -0.75 & $22-30$MeV \\
 \hline
\end{tabular}
\end{center}
\caption{ Input parameters. }
\label{input}
\end{table}

\begin{table}
\begin{center}
\begin{tabular}{|c|c|c|c|c|}
\hline  Quantity$\downarrow$ ~~~~$B\rightarrow$   & $N$ & $\Sigma$  & $\Xi$ & $\Lambda$ \\ \hline
$\bar u^B$ & 0.221  & 0.099  & 0.947 & 0.217  \\
$\bar d^B$ & 0.339  & 0.335  & 0.213 & 0.217 \\
$\bar s^B$ & 0.091  & 0.068  & 0.046 & 0.068  \\
$\bar u^B/\bar d^B$ & $0.652$  & $0.295$  & $0.445$ & 1  \\
$\bar u^B-\bar d^B$ & $-0.118$  & $-0.236$  & $-0.118$ & 0  \\
$f_u^B=\frac{u^B+\bar u^B}{\sum (u^B+\bar u^B)}$ & 0.567  & 0.549  & 0.321 & 0.358  \\
$f_d^B=\frac{d^B+\bar d^B}{\sum (d^B+\bar d^B)}$ & 0.390   & 0.167  & 0.115 & 0.358  \\
$f_s^B=\frac{s^B+\bar s^B}{\sum (s^B+\bar s^B)}$ &  0.042 & 0.283  & 0.564 & 0.284  \\
$f_0^B=f_u^B+f_d^B+f_s^B$ & 1  &1  &1 &1  \\
$f_3^B=f_u^B-f_d^B$ &  0.177 & 0.381  & 0.206 & 0  \\
$f_8^B=f_u^B+f_d^B-2f_s^B$ & 0.874   & 0.149  & $-0.693$ & 0.149  \\
$\frac{f_3^B}{f_8^B}$ & 0.203  & 2.563  &$-0.297$ & 0  \\
$\rho_s^B=\frac{ s^B+\bar{s}^B}{ u^B+ d^B}$ &  0.051 & 0.622  & 0.632 & 0.467  \\
$\kappa_s^B=\frac{  s^B+ \bar{s}^B}{\bar{u}^B+ \bar{d}^B}$ &  0.323 & 4.920  & 6.798 & 2.617  \\
$\frac{\sum \bar q^B}{\sum q^B}$ & 0.178  & 0.143   & 0.105  & 0.143   \\

\hline
\end{tabular}
\caption{ The $\chi$CQM results for the sea quark flavor distribution functions and related flavor dependent functions for the $N$, $\Sigma$, $\Xi$ and $\Lambda$ octet baryons.}\label{flavor-dist}
\end{center}
\end{table}

Using the above set of parameters, the results of our calculations pertaining to the the sea quark flavor distribution functions and related flavor dependent functions discussed above for the case of $N$, $\Sigma$, $\Xi$ and $\Lambda$ baryons, have been presented in Table \ref{flavor-dist}.
The present experimental situation, for the case of $N$, as obtained from the
DIS and Drell-Yan experiments \cite{nmc,baldit,e866} is given as follows
\bea
{\bar u^N-\bar d^N}_{NMC}&=& -0.147 \pm 0.024\,, \nonumber \\
{\bar u^N/\bar d^N}_{NA51}&=&0.51 \pm 0.09\,, \nonumber \\
{\bar u^N-\bar d^N}_{E866}&=& -0.118 \pm 0.018, \nonumber \\
{\bar u^N/\bar d^N}_{E866}&=& 0.67 \pm 0.06\,, \nonumber \\
{f_s^N}_{CCFR} &=& 0.076 \pm 0.02\,, \nonumber \\
{f^N_3/f^N_8}_{CCFR} &=& 0.21 \pm .05\,, \nonumber \\
{\rho_s^N}_{CCFR} &=& 0.099 \pm 0.009\,, \nonumber \\
{\kappa_s^N}_{CCFR} &=& 0.477 \pm 0.051\,, \nonumber \\
\frac{\sum \bar q^N}{\sum q^N} &=&0.245 \pm 0.005\,.
\label{antiquark-values}
\eea

The NQM, which is quite successful in explaining a good deal of low energy data \cite{Isgur,dgg,yaouanc},
has the following predictions for the above mentioned quantities
\bea
{\bar u^N-\bar d^N}&=& 0\,, \nonumber \\
{\bar u^N/\bar d^N}&=& -\,, \nonumber \\
f_s^N &=& 0\,, \nonumber \\
f^N_3/f^N_8 &=& \frac{1}{3}\,, \nonumber \\
\rho_s^N&=& 0\,, \nonumber \\
\kappa_s^N&=& 0\,, \nonumber \\
\frac{\sum \bar q}{\sum q} &=&0\,.
\label{antiquark-NQM}
\eea

From  Table \ref{flavor-dist} and Eqs. (\ref{antiquark-values}) and (\ref{antiquark-NQM}) we find that the important measurable quark distribution functions look to be in agreement with the most recent phenomenological/experimental results available which the NQM is unable to explain. For example, the $\chi$CQM results clearly indicate that the nucleon sea contains more number of $\bar d^N$ quarks than the $\bar u^N$ quarks as indicated by DIS and Drell-Yan experiments \cite{nmc,baldit,e866}. The $\chi$CQM result for ${\bar u^N/\bar d^N}$ is 0.652 and is clearly in agreement with the latest DIS results available for the case of nucleon ${\bar u^N/\bar d^N}= 0.67 \pm 0.06$ \cite{e866}. It is also quite in agreement with the results of Drell-Yan experiment giving ${\bar u^N/\bar d^N}=0.51 \pm 0.09$ \cite{baldit}. For the case of ${\bar u^N-\bar d^N}$, the $\chi$CQM gives $-0.118$ which is completely in agreement with the result of the latest Fermilab E866 experiment ${\bar u^N-\bar d^N}= -0.118 \pm 0.018$ \cite{e866}. The result of the earlier NMC experiment  is on the higher side ${\bar u^N-\bar d^N}= -0.147 \pm 0.024$ \cite{nmc}.

For the case of $f_s^N$, the NQM results show that this fraction of strange quarks is zero whereas the $\chi$CQM result comes out to be 0.042 which is close to the available data from CCFR ${f_s^N}= 0.076 \pm 0.02$ \cite{ccfr1,ccfr2}. Similarly, $\rho_s^N$ and $\kappa_s^N$ are predicted to be zero in NQM but $\chi$CQM predicts them to be 0.051 and 0.323 respectively. The results when compared with the available data ${\rho_s^N}= 0.099 \pm 0.009$ and ${\kappa_s^N}= 0.477 \pm 0.051$ \cite{ccfr1,ccfr2} clearly indicate that $\chi$CQM predict these quantities with the right magnitude and sign. Further, the ratio of total number of the antiquarks and quarks in $\chi$CQM for the case of nucleon is $\frac{\sum \bar q^N}{\sum q^N} =0.178$ as compared to the available phenomenological result $\frac{\sum \bar q^N}{\sum q^N} =0.245 \pm 0.005$. Our results for the quantities discussed above are also in agreement with the results predicted by other model calculations \cite{ellis-brodsky,alkofer,christov,diakonov,mesoncloud,wakamatsu}.

Since the understanding of the deep inelastic results as well as the dynamics of the constituents of the nucleon constitute a major challenge for any model trying to explain the nonperturbative regime of QCD, the success of $\chi$CQM not only justifies but also strengthens our conclusion regarding  the qualitative and quantitative role of the sea quarks in right direction. The non-vanishing values for strangeness dependent quantities for the case of nucleon indicate that the chiral symmetry breaking as well as SU(3) symmetry breaking are essential to understand the significant role played by the strange quarks in the nucleon.
Since no data is available for the $\Sigma$, $\Xi$ and $\Lambda$ octet baryons, any future measurement of these would
have important implications for the subtle features of $\chi$CQM.

\section{Flavor Structure functions and the Gottfried integral}

The basic flavor structure functions $F_1$ and $F_2$ are defined as \bea F_2^B(x) &=& x \sum_{u,d,s} e^2_{q}[q^B(x)
+ \bar q^B(x)] \,,\\ F_1^B(x) &=& \frac{1}{2x}F_2^B(x) \,. \eea
Using the quark distribution
functions from Eq. (\ref{totalquark}), the structure function $F_2$ for the baryons can be
expressed as
\bea  F^p_2(x) &=&\frac{4}{9} x\left(u^p_{{\rm V}}(x)+ 2\bar u^p(x)\right)
+\frac{1}{9} x\left(d^p_{{\rm V}}(x)+ 2 \bar d^p(x)+ s^p_{{\rm V}}(x))+2\bar s^p(x)\right)\,, \nonumber\\
  F^{\Sigma^+}_2(x) &=&\frac{4}{9} x\left(u^{\Sigma^+}_{{\rm V}}(x)+ 2\bar u^{\Sigma^+}(x)\right)
+\frac{1}{9} x\left(d^{\Sigma^+}_{{\rm V}}(x)+ 2 \bar d^{\Sigma^+}(x)+ s^{\Sigma^+}_{{\rm V}}(x))+2\bar s^{\Sigma^+}(x)\right)\,, \nonumber\\
 F^{\Xi^0}_2(x) &=& \frac{4}{9} x\left(u^{\Xi^0}_{{\rm V}}(x)+ 2\bar u^{\Xi^0}(x)\right)
+\frac{1}{9} x\left(d^{\Xi^0}_{{\rm V}}(x)+ 2 \bar d^{\Xi^0}(x)+ s^{\Xi^0}_{{\rm V}}(x))+2\bar s^{\Xi^0}(x)\right)\,, \nonumber\\
  F^{\Lambda^0}_2(x) &=&\frac{4}{9} x\left(u^{\Lambda^0}_{{\rm V}}(x)+ 2\bar u^{\Lambda^0}(x)\right)
+\frac{1}{9} x\left(d^{\Lambda^0}_{{\rm V}}(x)+ 2 \bar d^{\Lambda^0}(x)+ s^{\Lambda^0}_{{\rm V}}(x))+2\bar s^{\Lambda^0}(x)\right)\,.
 \eea

The deviation from the Gottfried sum rule \cite{gsr} can be obtained from the structure
functions of different isospin multiplets measured through the Gottfried integral $I^{B_1 B_2}_G$ for the octet baryons.
This experimentally observed quantity measures the
asymmetry between the $\bar u^B$ and the $\bar d^B$ quarks content in
the sea quarks. The Gottfried integrals can be simplified and expressed as follows
\bea I^{p n}_G \equiv \int_0^1
{\frac{F^p_2(x) -F^n_2(x)}{x}} dx &=& \frac{1}{3} + \frac{2}{3}
\left[ \bar u^p- \bar d^p \right]\,,\nonumber\\
I^{\Sigma^+ \Sigma^0}_G \equiv \int_0^1
\frac{F^{\Sigma^+}_2(x) -F^{\Sigma^0}_2(x)}{x} dx &=& \frac{1}{3}
+ \frac{2}{9} \left[ 4 \bar u^{\Sigma^+}+ \bar
d^{\Sigma^+} - 4 \bar u^{\Sigma^0} -\bar d^{\Sigma^0}
\right] \,,\nonumber\\ I^{\Sigma^0 \Sigma^-}_G \equiv \int_0^1
\frac{F^{\Sigma^0}_2(x) -F^{\Sigma^-}_2(x)}{x} dx &=& \frac{1}{3}
+ \frac{2}{9} \left[ 4 \bar u^{\Sigma^0}+ \bar
d^{\Sigma^0} - 4 \bar d^{\Sigma^-} - \bar u^{\Sigma^-}
\right] \,,\nonumber\\  I^{\Xi^{0} \Xi^{-}}_G \equiv \int_0^1{
\frac{F^{\Xi^0}_2(x) -F^{\Xi^-}_2(x)}{x}} dx &=& \frac{1}{3} +
\frac{2}{3}  \left[\bar u^{\Xi^0} - \bar d^{\Xi^0}
\right] \,. \label{ioctet}\eea
The normalization
conditions for the valence quarks used to derive the above equations have been taken from Table \ref{valencequark} whereas the sea quark contributions corresponding to each baryon obey the following normalization conditions  \[ \int_0^1 \bar
d^n(x)dx= \int_0^1 \bar u^p(x) dx \,,~~  \int_0^1 \bar u^n(x)dx =
\int_0^1 \bar d^p(x) dx \,,~~ \int_0^1 \bar s^n(x)dx= \int_0^1
\bar s^p(x) dx \,, \]
\be \int_0^1 \bar
u^{\Xi^-}(x)dx= \int_0^1 \bar d^{\Xi^0}(x) dx \,,~~  \int_0^1 \bar d^{\Xi^-}(x)dx =
\int_0^1 \bar u^{\Xi^0}(x) dx \,,~~ \int_0^1 \bar s^{\Xi^-}(x)dx= \int_0^1
\bar s^{\Xi^0}(x) dx \,. \ee

A measurement of the Gottfried integral for the
case of nucleon has shown a clear violation of Gottfried sum rule
from $\frac{1}{3}$  which can find its explanation in a global
quark sea asymmetry $ \int_0^1 (\bar d(x) -\bar u(x))dx$ which has been measured in the NMC and E866 experiments \cite{nmc,e866}. It is clear from Eq. (\ref{ioctet}) that the flavor symmetric sea ($\bar u^B$=$\bar d^B$) leads to $I_G=\frac{1}{3}$.
Similarly, for the case of $\Sigma^{+}$, $\Sigma^{0}$, and
$\Xi^0$, the Gottfried sum rules should read $I^{\Sigma^+
\Sigma^0}_G=\frac{1}{3}$, $I^{\Sigma^0 \Sigma^-}_G=\frac{1}{3}$
and $I^{\Xi^0 \Xi^-}_G=\frac{1}{3}$ for symmetric sea quarks. However, due to the $\bar d(x)-\bar u(x)$ asymmetry in
the case of octet baryons, a lower value of the Gottfried
integrals is obtained and the numerical values are given as follows
\bea I^{p n}_G &=& 0.254\,,\nonumber\\
I^{\Sigma^+ \Sigma^0}_G &=& 0.640 \,,\nonumber\\
I^{\Sigma^0 \Sigma^-}_G  &=& 0.569 \,,\nonumber\\
I^{\Xi^{0} \Xi^{-}}_G &=& 0.254 \,. \eea
For the case of nucleon, the $\chi$CQM result ($I^{p n}_G = 0.254$) is in good
agreement with the available experimental data of E866 \cite{e866}. We have $I^{p n}_G = \frac{1}{3} + \frac{2}{3}
\left[ \bar u^p- \bar d^p \right]=0.266 \pm 0.005$ from the NMC results \cite{nmc} and $I^{p n}_G = 0.254 \pm 0.005$ from the E866 results \cite{e866}.
Since no experimental results are available for the other octet baryons, new
experiments aimed at measuring the flavor content of the other
octet baryons are needed for profound understanding of the
nonperturbative properties of QCD as well as to understand the important role of the sea quarks at low value of $x$.

\section{meson-baryon sigma terms}
\label{pionsig}

The meson-baryon sigma term ($\sigma_{MB}$) corresponding to the pseudoscalar mesons and octet baryons is affected by the
contributions of the sea quark. It can be defined in terms of the scalar
quark content (($q {\bar q})_M$) of the particular meson $M$ ($\pi$, $K$ and $\eta$)
\bea
\sigma_{MB} = \hat{m} \langle B| (q {\bar q})_M |B
\rangle \,, \eea
where ${\hat m}$ is
the average value of current $u$ $d$ and $s$ quark masses evaluated at
fixed gauge coupling. For example, we have
\bea
\sigma_{\pi B} = \hat{m} \langle B| {\bar{u}u} + {\bar{d}d}|B
\rangle \,. \eea

The kaon-nucleon sigma term ($\sigma_{K B}$) can be expressed in
terms of the scalar quark content of $u$ and $d$ quarks as \be
\sigma_{K B} = \frac{\sigma^u_{K B} + \sigma^d_{K B}}{2} \,, \ee
where $$\sigma^u_{K B} = \frac{\hat{m} + m_s}{2}\langle B|
{\bar{u}u} + {\bar{s}s}|B\rangle \,,$$ and $$\sigma^d_{K B} =
\frac{\hat{m} + m_s}{2}\langle B| {\bar{d}d} +
{\bar{s}s}|B\rangle\,.$$ Similarly, the $\eta$-nucleon sigma term
($\sigma_{\eta B}$) can be expressed as \bea \sigma_{\eta B} &=&
\frac{1}{3} \langle B|\hat{m} ({\bar{u}u}+{\bar{d}d})+ 2 m_s
{\bar{s}s}|B\rangle \,. \eea The $\sigma_{K B}$ and $\sigma_{\eta
B}$ can be expressed in terms of the $\sigma_{\pi B}$ and $y_B$,
\bea \sigma_{K B} &=& \frac{\hat m + m_s}{2 \hat m}(1 +2 y_B)
\sigma_{\pi B} \,, \\ \sigma_{\eta B} &=& \frac{1}{3} \hat{\sigma}+
\frac{2 (m_s+ {\hat m})}{3 {\hat m }}y_B \sigma_{\pi B} \,,\eea
where we have defined \be
\hat {\sigma} = {\hat m}{\langle B| {\bar{u}u} + {\bar{d}d} -2
{\bar{s}s} | B \rangle}\,,\ee  and
\bea y_B &=& \frac{\langle B|{\bar{s}} s |B \rangle} {\langle B|
{\bar{u}u} + {\bar{d}d} | B \rangle}\,. \label{yb} \eea
In terms of $\hat {\sigma}$ and $y_B$ we can also define $\sigma_{\pi B}$ as
\be \sigma_{\pi B} =\frac{\hat {\sigma}}{1-2 y_N}. \label{sigmapiB}\ee

Another important parameter pertaining to the strangeness content in a baryon
is the strangeness sigma term \be
\sigma_s^B=m_s {\langle B| \bar{s}s | B \rangle} =\frac{1}{2}y_B
\frac{m_s}{\hat m} \sigma_{\pi B}\,. \ee
Using the respective antiquark flavor distribution
functions from the Table \ref{antiquark}, the meson-baryon sigma terms can be calculated and the results for the meson-baryon
sigma terms for $N$, $\Sigma$ and $\Xi$  have
been presented in Table \ref{baryon}. Since the $\sigma$ terms are characterized by
the light quark mass ratio $\frac{m_s}{\hat m}$, therefore, in addition to the parameters of $\chi$CQM listed in Table \ref{input}, we have used the most widely accepted range for $\frac{m_s}{\hat m}$ as  $22-30$MeV \cite{PDG}.
From Table \ref{baryon}, we find that the $\sigma$ terms are positive for the case of $N$ and $\Sigma$ however they are negative for the case of $\Xi$. This is clearly due to the dominance of the $s$ quarks in the valence structure of $\Xi$ due to which a higher value of $y_B$ (Eq. (\ref{yb})) is obtained. This leads to negative value of $\sigma_{\pi B}$ as defined in Eq. (\ref{sigmapiB}).

\begin{table}
\begin{center}
\begin{tabular}{|c|c|c|c|c|}
\hline  Quantity  & $N$ & $\Sigma$  & $\Xi$ & $\Lambda$ \\ \hline
$y_B$ & 0.044  & 0.396  & 1.294 & 0.396  \\
$\sigma_{\pi B}$ & 31.325  & 137.568  & $-17.974$ & 137.568 \\
$\sigma_{K B}$ & 195.952  & 1417.75  & $-370.992$ & 1417.75  \\
$\sigma_{\eta B}$ & $30.635$  & $845.167$  & $-347.328$ & $845.167$  \\
$\sigma^s_B$ & 15.145  & 599.483  & $-256.003$ & 599.483 \\

\hline
\end{tabular}
\end{center}
\caption{The $\chi$CQM results for the meson-baryon sigma terms
for the quark mass ratio $\frac{m_s}{\hat m}=22$.} \label{baryon}
\end{table}

The strangeness fraction of the nucleon from Eq. (\ref{fractionq}) can be related to the strangeness content from Eq. (\ref{yb})
as \be f_s^N = \frac{y_N}{1-y_N} \,, \label{fs} \ee which in terms of $\sigma_{\pi N}$  and $\hat {\sigma}$ can be
expressed as \bea f_s^N = \frac{\sigma_{\pi N} - \hat{\sigma}}{3 \sigma_{\pi N} -
\hat{\sigma}} \,.\eea According to NQM, the
valence quark structure of the nucleon does not involve strange
quarks. The validity of OZI rule \cite{ozi} in this case would
imply $y_N=f_s^N=0$ or ${\hat \sigma}=\sigma_{\pi N}$. For
$\frac{m_s}{\hat m}=22$, the value of $\sigma_{\pi N}$ comes out
to be close to 28 $MeV$. However, the most recent analysis of
experimental data gives higher values of $\sigma_{\pi N}$ which
points towards a significant strangeness content in the nucleon. The $\chi$CQM results giving a comparatively higher value of $\sigma_{\pi N}$ justify the mechanism of chiral symmetry breaking  and SU(3) symmetry breaking. Since no data is available for the $KN$ and $\eta N$ sigma terms as well for all the $MB$ terms corresponding to $\Sigma$ and $\Xi$ baryons, the future DA$\Phi$NE experiments \cite{dafne} for the determination of KN sigma
terms as well as information from the hyperon-antihyperon production in
heavy ion collisions will provide information of the contribution of the sea quark.
The results can however be compared with the other available phenomenological and theoretical results presented in Table \ref{data}.

\begin{table}
\begin{center}
{\begin{tabular}{|l| c|c|c|c|c|} \hline \TL{Phenomenological results }{$y_N$}{$\sigma_N^{s}$}{$\sigma_{\pi
N}$}{$\sigma_{K N}$} {$\sigma_{\eta N}$} \TL{of other theoretical
approaches }{}{}{(MeV)}{(MeV)}{(MeV)} \hline \TL{Koch {\it et al.}
\cite{koch}}{...}{...}{$64\pm8$}{...}{...} \TL{Gasser {\it et al.}
\cite{gasser}} {0.11 $\pm$ 0.07} {}{$45\pm8$}{...}{...}
\TL{Pavan {\it et al.} \cite{Pavan:2001wz}}{0.23}{...}{$79\pm
7$}{...}{...} \TL{Hite {\it et al.} \cite{hite}}{...}{...}{81
$\pm$ 6}{...}{...}
\TL{QCD Sum Rules \cite{qcdsr}}{0.17 $\pm$ 0.03}{161$\pm$
66}{53$\pm$24}{...}{...} \TL{Lattice QCD, Fukugita {\it et al.}
\cite{fukugita}}{0.33 $\pm$ 0.07}{...}{40 $-$ 60}{451 $\pm$
54}{...} \TL{Lattice QCD, Dong {\it et al.} \cite{dong}}{0.36
$\pm$ 0.03}{...}{$49.7 \pm 2.6$}{362$\pm$13}{...} 
\TL{Perturbative
quark model \cite{lyubovitskij}}
   {0.076 $\pm$ 0.012}{...}{$45\pm5$}{312 $\pm$ 37}{72 $\pm$ 16}
\TL{Heavy baryon chiral
perturbation theory \cite{borasoy2}}{0.12$\pm$0.03}{150 $\pm$
50}{45}{...}{...} 
\TL{
SU(3) Nambu-Jona-Lasinio Soliton model
\cite{kim}}{0.13}{...}{40.80}{...}{...} \TL{Chiral quark-soliton
model \cite{cqsm}}{...}{...}{67.9}{...}{...} \TL {Meson-cloud
model \cite{stuckey}}{...}{260}{45}{...}{...}  
 \hline \end{tabular} }
\end{center}
\caption{Phenomenological results of some other theoretical
approaches for strangeness content in the nucleon and
meson-nucleon sigma terms.} \label{data}
\end{table}

\section{summary and conclusions}

To summarize, the  quark flavor distribution functions of the octet baryons ($N$, $\Sigma$, $\Xi$, $\Lambda$) have been phenomenologically estimated in the
chiral constituent quark model ($\chi$CQM) since the understanding of the DIS results as well as the dynamics of the constituents of the baryon constitute a major challenge for any model trying to explain the nonperturbative regime of QCD. These quantities have important implications for the sea quark contributions, chiral symmetry breaking as well as SU(3) symmetry breaking. The valence and sea quark flavor distribution functions of the scalar density matrix elements of octet baryons have been computed explicitly for the $u$, $d$ and $s$ quarks in each baryon. To understand the role of sea quarks in understanding the important experimentally measurable quantities, the implications of this model have been studied for the sea quark asymmetries, fraction of a particular quark (antiquark) present in a baryon, flavor structure functions and the Gottfried integral. The $\chi$CQM results are in agreement with the most recent phenomenological/experimental results available and justifies the qualitative and quantitative role of the sea quarks in right direction. This can perhaps be substantiated further by a measurements for the other octet baryons.
The recent available experimental results pointing out a significant contribution of the strangeness in the nucleon also finds an answer in this model which gives a significant strangeness fraction of the nucleon.  The meson-baryon sigma terms $\sigma_{\pi B}$, $\sigma_{K B}$, and $\sigma_{\eta B}$ for the case of $N$, $\Sigma$ and $\Xi$ baryons have also been calculated. Since no data is available for the $\Sigma$ and $\Xi$ octet baryons, any future measurement of these would have important implications for the subtle features of $\chi$CQM. To conclude,  chiral symmetry breaking is the key to understand the contribution of the sea quarks in the nonperturbative regime of QCD where, at the leading order, the valence quarks and the weakly interacting Goldstone bosons constitute the appropriate degrees of freedom.

\section*{ACKNOWLEDGMENTS}

H. D. would like to thank Department of Science and Technology (Ref No. SB/S2/HEP-004/2013), Government of India, for financial support.

\end{document}